\documentclass{emulateapj}

\newcommand{\eg}{e.g., }
\newcommand{\ie}{i.e., }
\newcommand{\Msun}{M_{\odot}}
\newcommand{\Lsun}{L_{\odot}}
\newcommand{\Lsd}{L_{\rm sd}}
\newcommand{\Lp}{L_{\rm plateau}}
\newcommand{\tp}{t_{\rm plateau}}
\newcommand{\tausd}{\tau_{\rm sd}}
\newcommand{\Lsdi}{L_{\rm sd,0}}
\newcommand{\Menv}{M_{\rm env}}
\newcommand{\xh}{X_{\rm env}({\rm H})}
\newcommand{\kms}{km~s$^{-1}$}
\newcommand{\ergs}{erg~s$^{-1}$}

\newcommand{\Cofs}{$^{56}$Co}
\newcommand{\Nifs}{$^{56}$Ni}
\newcommand{\Mej}{M_{\rm ej}}
\newcommand{\Mms}{M_{\rm MS}}

\def\gsim{\mathrel{\rlap{\lower 4pt \hbox{\hskip 1pt $\sim$}}\raise 1pt
\hbox {$>$}}}
\def\lsim{\mathrel{\rlap{\lower 4pt \hbox{\hskip 1pt $\sim$}}\raise 1pt
\hbox {$<$}}}

\newcommand{\Mni}{M{\rm (^{56}Ni)}}

\slugcomment{Accepted for publication in the Astrophysical
 Journal Letters.}

\begin{document}

\title{Supernova Explosions of Super-Asymptotic Giant Branch Stars: Multicolor
Light Curves of Electron-Capture Supernovae}

\author{
 Nozomu~Tominaga\altaffilmark{1,2},
 Sergei~I.~Blinnikov\altaffilmark{3},
 Ken'ichi~Nomoto\altaffilmark{2,4}
 }
\altaffiltext{1}{Department of Physics, Faculty of Science and
Engineering, Konan University, 8-9-1 Okamoto,
Kobe, Hyogo 658-8501, Japan; tominaga@konan-u.ac.jp}
\altaffiltext{2}{Kavli Institute for the Physics and Mathematics of the
Universe (WPI), The University of Tokyo, 5-1-5 Kashiwanoha, Kashiwa, Chiba
277-8583, Japan}
\altaffiltext{3}{Institute for Theoretical and  Experimental Physics (ITEP),
Moscow 117218, Russia; Sergei.Blinnikov@itep.ru}
\altaffiltext{4}{Department of Astronomy, School of Science,
The University of Tokyo, Bunkyo-ku, Tokyo 113-0033, Japan;
nomoto@astron.s.u-tokyo.ac.jp}

\setcounter{footnote}{4}

\begin{abstract}
 An electron-capture supernova (ECSN) is a core-collapse supernova (CCSN)
 explosion of a super-asymptotic giant branch (SAGB) star with a main-sequence
 mass $\Mms\sim7-9.5\Msun$. The explosion takes place in accordance with
 core bounce and subsequent neutrino heating and is a unique example 
 successfully produced by first-principle simulations. This allows us to
 derive a first self-consistent
 multicolor light curve of a CCSN. Adopting the
 explosion properties derived by the first-principle
 simulation, \ie the low explosion energy of $1.5\times10^{50}$~erg and the
 small \Nifs\ mass of $2.5\times10^{-3}\Msun$, we perform a multigroup
 radiation hydrodynamics calculation of ECSNe
 and present multicolor light curves of ECSNe of SAGB stars with various
 envelope mass and hydrogen abundance.
 We demonstrate that a shock breakout has a peak luminosity of
 $L\sim2\times10^{44}$~\ergs\ and can evaporate circumstellar dust up to
 $R\sim10^{17}$~cm for a case of carbon dust, that plateau
 luminosity and plateau duration of ECSNe are $L\sim10^{42}$~\ergs\ and
 $t\sim60-100$~days, respectively, and that a plateau is
 followed by a tail with a luminosity drop by $\sim4$~mag. 
 The ECSN shows a bright and short plateau that is as bright as typical
 Type II plateau supernovae, and a faint tail that might be influenced by
 spin-down luminosity of a newborn pulsar. Furthermore, the
 theoretical models are compared with ECSN candidates: SN~1054
 and SN~2008S. We find that SN~1054 shares the characteristics of the
 ECSNe. For SN~2008S, we find that its faint plateau requires an ECSN model with a
 significantly low explosion energy of $E\sim10^{48}$~erg.
\end{abstract}


\section{INTRODUCTION}
\label{sec:intro}

A massive star with a main-sequence mass $\Mms\gsim8\Msun$
ends up as a core-collapse supernova (CCSN). Core
collapse is inaugurated by electron capture for a star with an O+Ne+Mg core
($\Mms\lsim10\Msun$) or Fe photodisintegration for a star with an Fe core
($\Mms\gsim10\Msun$). 

An explosion mechanism of CCSNe is still under investigation, in
particular, for the CCSN of an star with an Fe core (Fe CCSN). 
Recently, sophisticated multidimensional simulations discovered that neutrino-driven convection
and/or standing-accretion shock instability enhances neutrino
heating and initiates an outward
flow \citep[\eg][for recent reviews]{jan12,kot12,bur13,bru13}. However,
explosion energies $E$ are about one orders of magnitude smaller than a
canonical value of a normal CCSN \citep[$E\sim10^{51}$~erg, \eg
SN~1987A][]{bli00}.

The fate of the less-massive star
with the O+Ne+Mg core is different from that of the star with an Fe
core \citep{miy80,nom82crab,nom84,nom87,miy87}. The O+Ne+Mg core is supported by electron
degenerate pressure. The mass and density of the O+Ne+Mg core increase
through phases of shell burning of He and H. As the O+Ne+Mg core
grows, an envelope undergoes mass loss to reduce the H mass and He
dredge-up to enhance He abundance. Eventually, the
star becomes a super-asymptotic giant branch (SAGB) star
\citep[\eg][]{sie07}. When the central density exceeds a critical value
($4\times10^{12}~{\rm kg~m^{-3}}$), electrons begin to be captured by
magnesium, the degenerate pressure decreases, and thus the O+Ne+Mg core
collapses gravitationally \citep{miy80,nom82crab}.

Ensuing core bounce and neutrino heating can eject the envelope and
part of the O+Ne+Mg core because of weak inward momentum carried by
the low-density SAGB envelope \citep{kit06,bur07,jan08}. The explosion is called an
electron-capture supernova \citep[ECSN;][]{nom84,nom87}. 
In contrast to the Fe CCSN, the explosion of the ECSN is
realized by first-principle simulations \citep{kit06,bur07,jan08} and 
a two-dimensional simulation demonstrates that the explosion takes place almost
spherically \citep{jan08}. This is the only example of
the CCSN being self-consistently produced. However, a low explosion
energy derived from the simulations ($E\sim10^{50}$~erg)
discriminates the ECSN from the normal CCSNe.

The first-principle hydrodynamics simulation enables self-consistent
studies, \eg\ on nucleosynthesis and observational features.
Explosive nucleosynthesis in the ECSN is calculated by \cite{hof08} and
\cite{wanajo09} based on \cite{kit06}. They find a small amount of
\Nifs\ in the ejecta [$\Mni\sim0.003\Msun$] and large production of elements with
$Z=30-40$. The small \Nifs\ mass is also decidedly different from the
normal CCSNe with $\Mni\sim0.07\Msun$. However, a theoretical
light curve of the ECSN has not been presented and the observational
features of the ECSN are not yet theoretically clarified.

There are several SNe which are suggested to be the ECSNe. One of the old
examples is the Crab Nebula being a remnant of SN~1054. This is suggested from
high He abundance \citep{mac08}, small ejecta mass
\citep{fes97}, and low kinetic energy \citep{fra95}. One of the recent
examples is SN~2008S \citep{pri08,bot09}, which is suggested from a
dust-surrounding bright progenitor and its faintness and slow
evolution. Other SN~2008S-like objects have also been discovered
\citep[\eg][]{bond09,szc12}.

In this Letter, we adopt the explosion properties derived by the
state-of-the-art first-principle simulation and present the first
self-consistent multicolor light curves of ECSNe. We also investigate
the contribution from a continuous energy release from a central remnant
as the Crab pulsar. Finally, we compare the multicolor light curves with
observations of ECSN candidates.

\begin{figure}[t]
\epsscale{1.}
\plotone{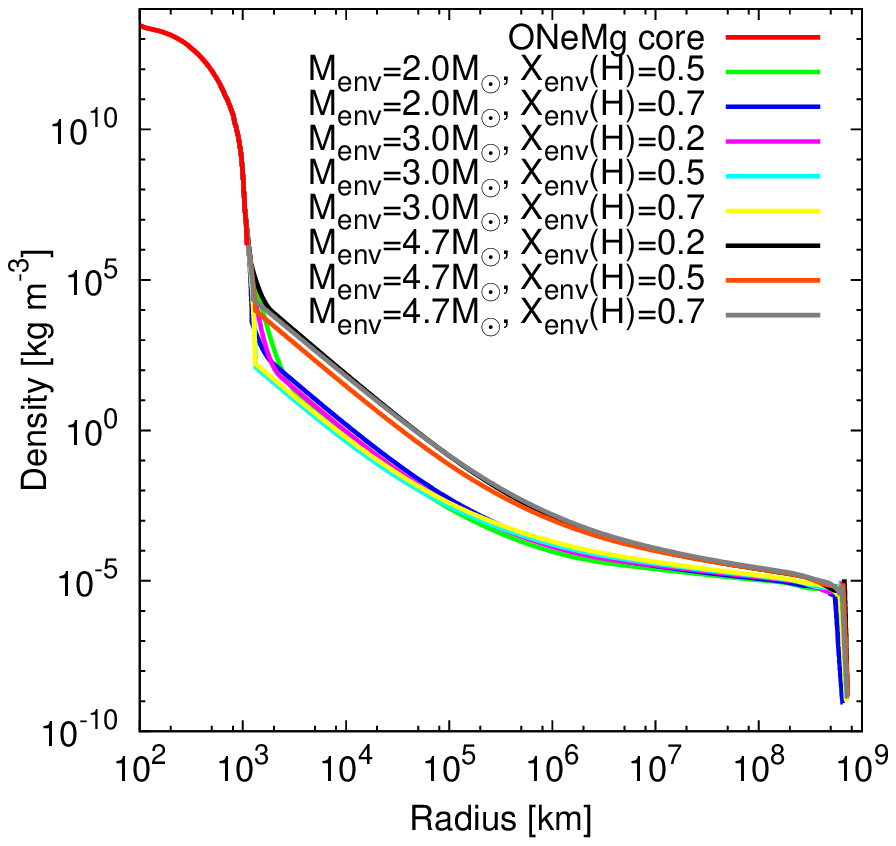}
\figcaption{Presupernova structures of SAGB stars with $\Menv=2.0\Msun$
 and $\xh=0.5$ (green), $\Menv=2.0\Msun$
 and $\xh=0.7$ (blue), $\Menv=3.0\Msun$
 and $\xh=0.2$ (magenta), $\Menv=3.0\Msun$
 and $\xh=0.5$ (cyan), $\Menv=3.0\Msun$
 and $\xh=0.7$ (yellow), $\Menv=4.7\Msun$
 and $\xh=0.2$ (black), $\Menv=4.7\Msun$
 and $\xh=0.5$ (orange), and $\Menv=2.0\Msun$
 and $\xh=0.7$ (gray) sitting on the $1.377\Msun$ O+Ne+Mg core (red). \label{fig:preSN}}
\end{figure}

\section{Model}
\label{sec:model}

\begin{figure}[t]
\epsscale{1.}
\plotone{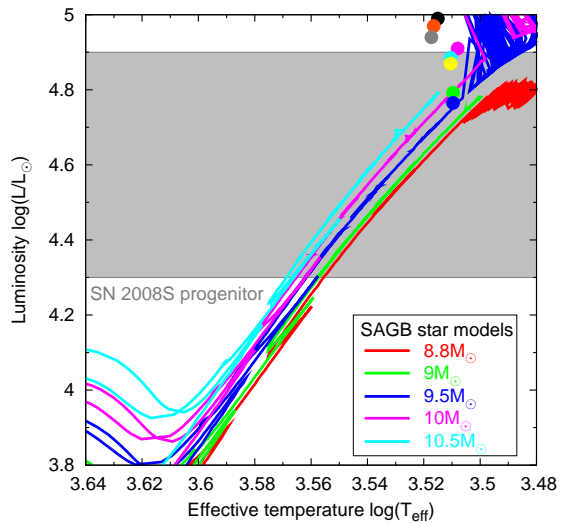}
\figcaption{Color-magnitude diagram of SAGB stars. The adopted progenitor
 models (filled circles; colors are the same as
 Fig.~\ref{fig:preSN}) is compared with 
 evolutionary models of SAGB stars with various $\Mms$ [lines,
 $\Mms=8.8\Msun$ (red), $9\Msun$ (green),
 $9.5\Msun$ (blue), $10\Msun$ (magenta), and $10.5\Msun$
 (cyan), \citealt{sie07}] and the
 luminosity range of the progenitor of SN~2008S (shaded gray 
 region, \citealt{pri08,bot09}). $\Lsun$ is the
 solar luminosity. \label{fig:HR}}
\end{figure}

We take an O+Ne+Mg core model with 1.377$\Msun$ at a 
presupernova stage from \cite{nom82crab} and \cite{nom84,nom87}. The
model is a core of a star with $\Mms=8.8\Msun$ which is calculated from
an He star with 2.2$\Msun$. A mass range of stars with the O+Ne+Mg core
is $\Mms\sim7-9.5\Msun$ but a progenitor of the ECSN should possess
an SAGB envelope \citep[see][for a review]{lan12}, of which mass and
abundance are influenced by $\Mms$, mass loss, and third dredge-up
associated with thermal pulses. However, the mass loss is
highly uncertain and no calculation of full thermal pulses has been
available. For almost fully convective envelope models of the
progenitor, therefore, we adopt various envelope
mass $\Menv$ ($=2.0-4.7\Msun$) and hydrogen abundance $\xh$ ($=0.2-0.7$)
by constructing hydrostatic and thermal equilibrium envelopes
with binding energies of $<10^{47}$~erg (\eg
\citealt{sai88}). Density structures are shown in
Figure~\ref{fig:preSN}. The luminosity of the
progenitor models is $L\sim3\times10^{38}$~\ergs \citep{nom87} and
their radii are $R\sim7\times10^{8}$~km. Figure~\ref{fig:HR}
demonstrates that the luminosity is roughly consistent with the
progenitor of SN~2008S \citep{pri08,bot09} and is located at a bright
tip of the SAGB stars \citep[][]{sie07}.

The explosion is initiated by a thermal bomb\footnote{This
does not affect the result because thermal energy is
efficiently converted to kinetic energy by the steep density
gradient until a shock emerges from a stellar surface
(\S~\ref{sec:result}).} with the explosion energy derived
by the first-principle simulation ($E=1.5\times10^{50}$~erg,
\eg \citealt{kit06}). The subsequent evolution is followed by a
multigroup radiation hydrodynamical code {\sc stella}
\citep{bli98,bli00,bli06}, in which one-group $\gamma$-ray transfer
is calculated and in situ absorption of positron is assumed for energy
deposition from \Nifs-\Cofs\ radioactive decay. An abundance distribution in the O+Ne+Mg
core after the explosion and mass of heavy elements are taken from the Model
ST in \cite{wanajo09}, which yields $2.5\times10^{-3}\Msun$ of
\Nifs. We note that no \Nifs\ is synthesized in the envelope due to
the low temperature of $<2\times10^9$~K.

We also investigate the contribution from the pulsar spin-down luminosity
that could be bright at birth. For example, the initial spin-down
luminosity of the Crab pulsar was $3.3\times10^{39}$~\ergs.\footnote{This is
estimated from current spin-down luminosity
($\Lsd\sim5\times10^{38}$~\ergs, \citealt{hes08}) with an equation
$\Lsd(t)=\Lsdi/(1+t/\tausd)^{(n+1)/(n-1)}$, where $\Lsdi$,
$\tausd(=700~{\rm yr})$ and $n(=2.5)$ are initial spin-down
luminosity, a spin-down timescale, and a braking index of the Crab
pulsar, respectively.} Since the pulsar spin-down luminosity can vary
on individual ECSNe and the deposition efficiency of an energy
released from the pulsar is unknown, we expediently
adopt the initial spin-down luminosity of the Crab pulsar and assume that
the released energy is fully deposited at the bottom of the ejecta
(``full deposition'') or
deposited pursuant to the same one-group transport as $\gamma$-rays from
the radioactive decay (``one-group transport'').

\section{Evolution of Electron-Capture Supernovae}
\label{sec:result}

\begin{figure}[t]
\epsscale{1.}
\plotone{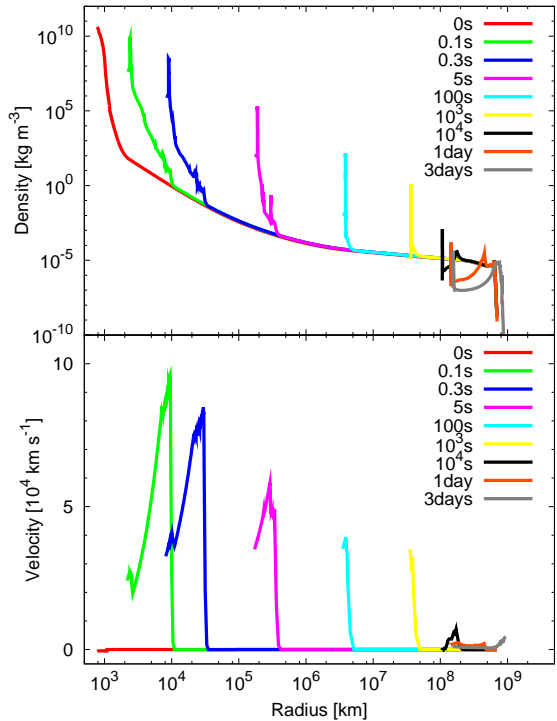}
\figcaption{Density (a) and velocity (b) structures of the ECSN of the
 SAGB star with $\Menv=3.0\Msun$ and $\xh=0.2$ at 0~s (red),
 0.1~s (green), 0.3~s (blue), 5~s (magenta), 100~s
 (cyan), 10$^3$~s
 (yellow), 10$^4$~s (black), $1$~day (orange), and
 $3$~days (gray)
 after the core collapse. \label{fig:str}
}
\end{figure}

In the explosion of the model with
$\Menv=3.0\Msun$ and $\xh=0.2$ (Figs.~\ref{fig:str}a-\ref{fig:str}b), a shock wave is
accelerated up to $9.6\times10^{4}$~\kms\ during the first $0.1$~s due
to the steep density gradient at the bottom of the envelope and
decelerated down to $1.4\times10^{3}$~\kms\ by shock emergence from
the stellar surface. Such
drastic deceleration of the shock develops severe Rayleigh-Taylor
instability.

\begin{figure}[t]
\epsscale{1.}
\plotone{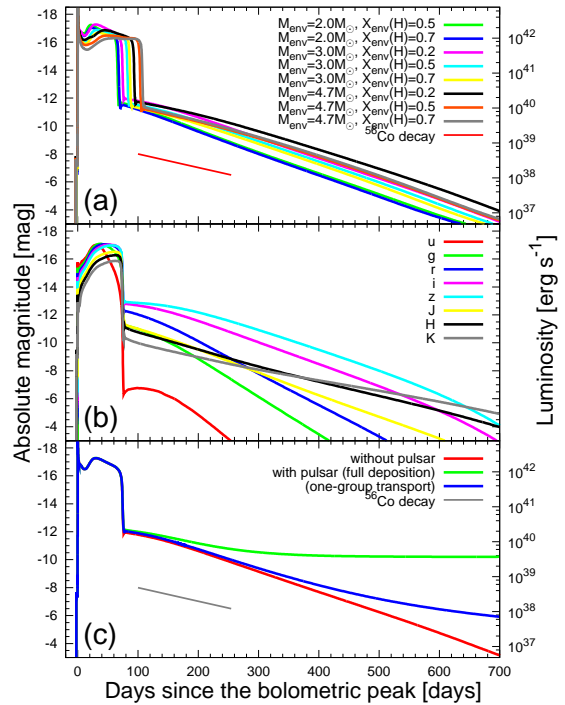}
\figcaption{(a) Bolometric light curves of the ECSNe. The
 colors are the same as Figure~\ref{fig:preSN} but the red line shows
 the energy release rate of \Cofs\ radioactive decay. (b) Multicolor
 light curves (red: $u$, green: $g$, blue: $r$, magenta: $i$, cyan: $z$,
 yellow: $J$, black: $H$, and gray: $K$) of the ECSN of the SAGB star
 with $\Menv=3.0\Msun$ and $\xh=0.2$. (c) Bolometric light curves of the
 ECSNe without the pulsar contribution (red), with the full
 deposition (green), and with the one-group
 transport (blue). The energy release
 rate of \Cofs\ radioactive decay (gray) is also shown. \label{fig:LCdep}}
\end{figure}

At the shock emergence, the ECSN flashes 
(Fig.~\ref{fig:LCdep}a). The phenomenon is
called a shock breakout, bolometrically brightest. Hereafter, we set $t=0$
at the shock breakout. A peak wavelength at the shock breakout is
$\sim200$~\AA\ and bolometric luminosity is as bright as
$L=2.4\times10^{44}$~\ergs. Adopting a dust evaporation radius
\citep{dwe83} and evaporation temperature for a short flash
\citep{pea86}, the shock breakout can destroy the circumstellar dust up to
$\sim9.6\times10^{11}$~km for a case of carbon dust.

As the SN ejecta cools down, the ECSN enters a plateau phase lasting
$\tp\sim60-100$~days. The plateau luminosity is
$\Lp\sim10^{42}$~\ergs, which is as bright as 
typical Type II plateau supernovae (SNe II-P, \eg SN~2004et,
\citealt{sah06}), and a photospheric velocity at the plateau is 
$3000-4000$~\kms, which is slightly slower than typical SNe II-P. The
plateau luminosity is fainter and the
duration is longer for an explosion of a star with larger $\Menv$ and
higher $\xh$.  The plateau is followed by a tail. The
luminosity suddenly fades by $\sim4$ mag at the transition because of the small
$\Mni$. The tail luminosity declines gradually at a rate of $0.012-0.016$~mag~day~$^{-1}$
which is slightly faster than the energy release rate from the \Cofs\
decay.

The multicolor light curves of the model with $\Menv=3.0\Msun$ and
$\xh=0.2$ are shown in Figure~\ref{fig:LCdep}b. The photospheric
temperature decreases after the shock breakout. Thus, the peak wavelength
shifts to redder bands with time and the optical luminosity increases
gradually. The light curves in the bluer bands peak at earlier epochs, \eg
$t\sim10$~days in the $u$ band and $t\sim20$~days in the $g$ band. In contrast
to typical SNe II-P, the $u$ and $g$ light curves do not
show a plateau and begin to decline immediately after the peak.
The light curves in the bands redder than the $i$ band peak at the end of the
plateau. The multicolor light curves drop at the transition to
the tail as the bolometric light curve does. The decline rates of
the tail are higher for bluer bands.

The pulsar spin-down only influences the tail
because the spin-down luminosity of the Crab pulsar is much fainter than the plateau
luminosity. The ECSN shines with the energy release from the \Cofs\
decay at the beginning of the tail and then the energy
release from the newborn pulsar becomes dominant at a late phase
($t\gsim250$~days, Fig.~\ref{fig:LCdep}c). The tail luminosity is
floored in the full deposition model, while the decline rate is
lowered in the one-group transport model. The tail luminosity
can be $\sim2-6$~mag brighter than the model without the pulsar contribution
at $t=600$~days.

\section{Comparison with the observations}
\label{sec:obs}

\subsection{Crab Nebula and SN~1054}

The Crab Nebula is the most famous and
conspicuous supernova remnant. Optical and UV observations
illustrate that the Crab Nebula is notably He-rich [$X({\rm He})\sim0.6-0.9$, 
\eg \citealt{mac08}]. In addition to the He-rich
abundance, a small ejecta mass ($\Mej=4.6\pm1.8\Msun$,
\citealt{fes97}) and low kinetic energy
($E<3\times10^{49}$~erg,\footnote{Here, $\Mej=1-2\Msun$ is
assumed. The kinetic energy is doubled by adopting
$\Mej=4.6\pm1.8\Msun$.} \eg \citealt{fra95}) suggest that the Crab
Nebula is a remnant of the ECSN.

An expansion of the Crab Nebula was discovered in the early 1920s
\citep{dun21} and, in the same year, the proximity of the Crab Nebula to
SN~1054 was
indicated \citep{lun21}. An explosion date estimated by turning back
the expansion is consistent with SN~1054 \citep{rud08}. Hence, it is 
widely believed that the expanding Crab Nebula is a remnant of SN~1054.

A sudden appearance of SN~1054 was recorded in medieval
times and its optical light curve is enscrolled in
historiographies
\citep{psk85,ste02}. They described the dates of the first and last
sightings and its brightness. However, the medieval observations were
rough and unconfident. Thus, reliability of the
archives is deeply scrutinized by \cite{ste02}. Referring the
conclusion of that paper, we adopt the following three points with errorbars of
$1$~mag and $20$~days; (1) SN~1054 was as bright as Venus (optical
magnitude $m_{\rm opt}\sim-3.5$ to $-5$) on July 4, 1054 and might be visible
earlier than July 4, 1054 (\eg May 10, 1054), (2) SN~1054 was visible in
daytime for 23 days from July 4, 1054 and $m_{\rm opt}\sim-3$ on July 27,
1054, and (3) SN~1054 disappeared in night on April 6, 1056 with
$m_{\rm opt}\sim6$. We correct the distance ($D=2$~kpc,
\citealt{tri73}) and reddening [$E(B-V)=0.52$, \citealt{mil73}] to the
Crab Nebula.

\begin{figure}[t]
\epsscale{1.}
\plotone{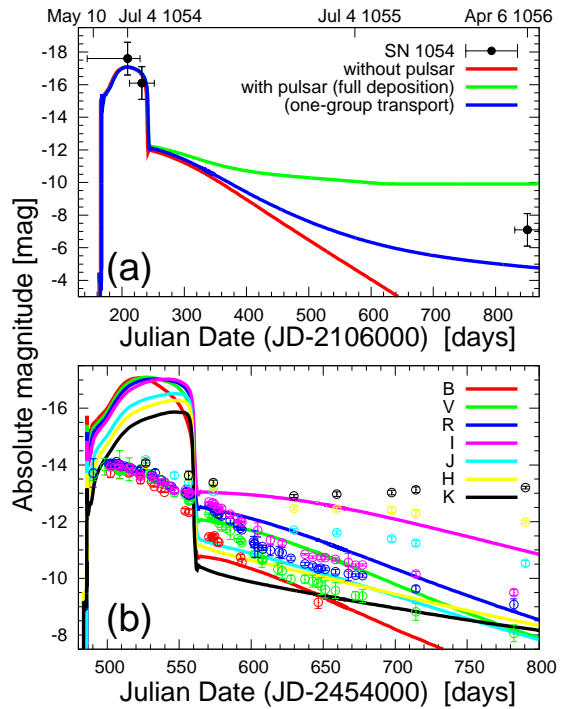}
\figcaption{(a) Optical light curves of SN~1054 (black circles) and
 the ECSNe of the SAGB stars. The colors are the
 same as Fig.~\ref{fig:LCdep}c. (b) Multicolor light curves of SN~2008S
 (red: $B$, green: $V$, blue: $R$, magenta: $I$, cyan: $J$,
 yellow: $H$, and black: $K$) are compared with those of the ECSN of the
 SAGB star with $\Menv=3.0\Msun$ and $\xh=0.2$.
\label{fig:LCcomp}}
\end{figure}

We compare an ``optical''\footnote{An optical band-pass is assumed to be a
normal distribution with a peak at $5550$~\AA\ and a
root-mean-square deviation of $550$~\AA\ \citep{vos78}.} light curve of
the model with $\Menv=3.0\Msun$ and $\xh=0.2$ against the observations of
SN~1054 (Fig.~\ref{fig:LCcomp}a). Here, an optical peak of the model is
set to be July 4, 1054. The ECSN model well reproduces the bright and
short plateau of SN~1054. The epoch of the transition from the plateau
to the tail corresponds to the date at which SN~1054
disappeared from the daytime sky. 

The tail luminosities of the full deposition and
one-group transport models are brighter and fainter than SN~1054,
respectively. If the deposition efficiency is the middle of these models,
a gradually-declining optical light curve can cross with the last
observation of SN~1054. We note that the last point of SN~1054
is not compatible with the ECSN model without the pulsar contribution as
suggested in \cite{sol01}.

\subsection{SN~2008S}

SN~2008S is suggested to be the ECSN because the progenitor is bright and
surrounded by dust and because SN~2008S is faint and evolves slowly
\citep{bot09}. However, the plateau luminosity of the ECSN is as bright as
that of normal SNe II-P due to the large
presupernova radius (Fig.~\ref{fig:LCdep}a). Thus, we conclude that the
ECSN model based on the first-principle simulation \citep{kit06} is incompatible with
the faintness of SN~2008S (Fig.~\ref{fig:LCcomp}b).

However, there is a caveat that the plateau luminosity and duration
depend on $E$ and $\Menv$ ($\Lp\propto E^{5/6} \Menv^{-1/2}$ and
$\tp\propto E^{-1/6} \Menv^{1/2}$, \citealt{lit85,pop93,eas94}). If an
explosion energy may be different for individual ECSNe, \eg due to
rotation, the faint and slowly-evolving SN~2008S could be
explained. Applying the scaling laws and taking
$\Lp\sim1\times10^{41}$~\ergs\ and $\tp\sim140$~days of SN~2008S,
we can derive its explosion properties;
$E\sim3.5\times10^{48}$~erg and $\Menv\sim3.4\Msun$. A series of
first-principle simulation and light curve calculations are required to
confirm that such the explosion is feasible and explains the multicolor
light curves of SN~2008S.

\section{Discussion and Conclusion}
\label{sec:discuss}

We present multicolor light curves of ECSNe with various $\Menv$
and $\xh$ based on the results of the 
first-principle simulation \citep{kit06}. We demonstrate that the shock
breakout has a peak luminosity of $L\sim2\times10^{44}$~\ergs\ and can evaporate
circumstellar dust up to $R\sim10^{12}$~km for the case of carbon dust and
that the plateau luminosity and the duration of the ECSNe are
$\Lp\sim10^{42}$~\ergs\ and $\tp\sim60-100$~days, respectively. The brighter
and shorter plateau is realized by the model with smaller $\Menv$ and
lower $\xh$. The plateau is followed by the tail with the
luminosity drop by $\sim4$~mag. 

The tail luminosity declines by $0.012-0.016$~mag~day~$^{-1}$ for the model
without the pulsar contribution, while, if the pulsar contributes to the
light curve, the tail light curve is floored or the decline rate is
lowered. The contribution from the pulsar spin-down luminosity as bright
as the Crab pulsar is prominent only in the tail. 

Furthermore, we compare the theoretical models with the ECSN candidates:
SN~1054 and SN~2008S. 

The bright and short plateau and low explosion energy of the ECSN model are
consistent with SN~1054. The plateau is not reproduced with a low-energy
explosion of a red supergiant star with heavy $\Menv$. The tail
luminosity of SN~1054 could be explained by the spin-down luminosity of
the newborn Crab pulsar. The deceleration of the shock wave in the ECSN
(Figs.~\ref{fig:str}a-\ref{fig:str}b) is also favorable to produce
filamentary structures observed in the Crab Nebula (\eg
\citealt{fes97}).\footnote{The filamentary structures can also be
originated by interaction between the ECSN ejecta and a pulsar wind
nebula (\citealt{hes08} for a review). However, the structures
exist even in the ejecta unaffected by the pulsar
\citep[\eg][]{rud08}. Future 3D calculations will test how the shock
deceleration produces the filamentary structures.} Thus the observed
features of SN 1054 are naturally reproduced by the ECSN model. 
The luminosity of interaction of the ECSN with a circumstellar medium
reaches only $\sim 10^{-3}$ of the plateau of SN 1054 for a typical SAGB
star with the mass loss rate of $\sim10^{-4}\Msun~{\rm yr^{-1}}$
\citep{smi13}.  Thus an extremely dense and confined ($r <$ a few
$10^{15}$~cm) circumstellar matter is required to explain the bright and
short plateau of SN 1054 \citep{smi13}.

The optical light curve of SN~1054 constrains the initial spin-down
luminosity of the Crab pulsar. Radio observations suggest $\tausd\leq30$~yr
and $\Lsdi\sim1.5\times10^{42}$~\ergs\ to produce relic
relativistic electrons \citep{ato99}. However, such high $\Lsdi$ leads to much
brighter optical luminosity than SN~1054 at April 6, 1056. Our result
favors $\tausd\sim700$~yr and $\Lsdi\sim3.3\times10^{39}$~\ergs. These
are also supported by a broad spectral evolution model of the pulsar
wind nebula \citep{tan10}.

The typical ejecta velocity of the ECSN model with
$\Menv=3.0\Msun$ and $\xh=0.2$ is
$v\sim2.2\times10^3$~\kms, which is consistent
with the low expansion velocity of the Crab Nebula \citep{rud08}.
On the other hand, the velocity of a wind blown from the progenitor is
$29$~\kms. Adopting the duration of the SAGB phase ($\sim10^4$~yr,
\citealt{nom82crab,sie07}), the SAGB wind extends
only upto $9.1\times10^{12}$~km ($=0.30$~pc), which is similar to an
apparent size of the Crab Nebula (\eg \citealt{rud08}). The
radius is smaller than the typical ejecta location of 
$6.9\times10^{13}$~km ($=2.2$~pc). Therefore, the forward shock should
locate in a low-density
circumstellar wind blown at the core He-burning phase or in an interstellar
medium. This could be a reason why the forward shock of the Crab Nebula
has not been found.

On the other hand, the plateau of the ECSN model is much brighter than that
of SN~2008S. If the explosion energy of ECSNe is exactly
$1.5\times10^{50}$~erg as derived by the first-principle simulation \citep{kit06}, the
multicolor light curves of the ECSN are inconsistent with those of
SN~2008S. However, there could
be a caveat that the explosion energy may vary on individual ECSNe. The
ECSN explosion with $E\sim3.5\times10^{48}$~erg and $\Menv\sim3.4\Msun$
might be compatible with SN2008S.

If SN~2008S is the ECSN with $E\sim3.5\times10^{48}$~erg and
$\Menv\sim3.4\Msun$, we can speculate on shock breakout luminosity
and $\Mni$; according to analytic dependence \citep{mat99}, the
luminosity of the shock breakout is $\sim1.4\times10^{42}$~\ergs.
Carbon dust at $\lsim9.2\times10^{10}$~km is evaporated by the shock
breakout. The size of a dust-free cavity is roughly consistent with
that estimated from a midinfrared observation of SN~2008S \citep{bot09}.
The explosion energy is comparable to a gravitational binding energy of the
progenitor at $M\geq1.3758\Msun$. Thus, SN~2008S is likely to eject
materials above the outer edge of the core and yield
$\Mni\sim4.4\times10^{-4}\Msun$, which is slightly smaller than an
estimate from the observed tail of SN~2008S
(\citealt{bot09}; but see also \citealt{koc11}). Therefore, we propose that the light curve tail of SN~2008S 
is powered by a spin-down luminosity of a newborn neutron star as
SN~1054. We note a caveat that SN~2008S is a Type IIn SN and thus
could be contributed to by interacting with a circumstellar medium.

The number of SNe sharing the characteristics of the ECSN, \ie a bright
and short plateau and faint tail, is small, if the SN~2008S-like
transients are not the ECSNe. We also note that faint and
low-energy SNe II-P, \eg SN~1997D and SN~1999br
\citep{zam03}, have a slower photospheric velocity at the plateau
($v\sim1000-2000$~\kms) than the ECSN. The scarcity
of the ECSNe might stem from a small mass range of a star ending up as ECSNe at
solar metallicity \citep[see][for a review]{lan12}. Since a significant 
contribution of ECSNe to the production of $^{48}$Ca, $^{64}$Zn, and $^{90}$Zr
is suggested \citep{wanajo09,wanajo11,wanajo13}, chemical evolution
models taking into account the scarcity is required to study the
origin of these isotopes.

\acknowledgments

We would like to thank Shuta J. Tanaka and Keiichi Maeda for fruitful
discussions on the Crab Pulsar and SNe II-P, respectively, and Shinya Wanajo
for providing the abundance distribution.
Data analysis were in part carried out on the general-purpose PC farm
at Center for Computational Astrophysics, National Astronomical
Observatory of Japan. 
This research has been supported in part by the
RFBR-JSPS bilateral program, World Premier
International Research Center Initiative, MEXT, Japan, and by the
Grant-in-Aid for Scientific Research of the JSPS (23224004, 23540262, 23740157)
and MEXT (23105705). S.B. is supported by 
the Grants of the Government of the Russian Federation 11.G34.31.0047,
RF Sci.~School 5440.2012.2 and 3205.2012.2 and by a grant
IZ73Z0-128180/1 of the Swiss National Science Foundation (SCOPES).


\clearpage

\begin{thebibliography}{51}
\expandafter\ifx\csname natexlab\endcsname\relax\def\natexlab#1{#1}\fi

\bibitem[{{Atoyan}(1999)}]{ato99}
{Atoyan}, A.~M. 1999, \aap, 346, L49

\bibitem[{{Blinnikov} {et~al.}(2000){Blinnikov}, {Lundqvist}, {Bartunov},
  {Nomoto}, \& {Iwamoto}}]{bli00}
{Blinnikov}, S., {Lundqvist}, P., {Bartunov}, O., {Nomoto}, K., \& {Iwamoto},
  K. 2000, \apj, 532, 1132

\bibitem[{{Blinnikov} {et~al.}(1998){Blinnikov}, {Eastman}, {Bartunov},
  {Popolitov}, \& {Woosley}}]{bli98}
{Blinnikov}, S.~I., {Eastman}, R., {Bartunov}, O.~S., {Popolitov}, V.~A., \&
  {Woosley}, S.~E. 1998, \apj, 496, 454

\bibitem[{{Blinnikov} {et~al.}(2006){Blinnikov}, {R{\"o}pke}, {Sorokina},
  {Gieseler}, {Reinecke}, {Travaglio}, {Hillebrandt}, \& {Stritzinger}}]{bli06}
{Blinnikov}, S.~I., {R{\"o}pke}, F.~K., {Sorokina}, E.~I., {Gieseler}, M.,
  {Reinecke}, M., {Travaglio}, C., {Hillebrandt}, W., \& {Stritzinger}, M.
  2006, \aap, 453, 229

\bibitem[{{Bond} {et~al.}(2009){Bond}, {Bedin}, {Bonanos}, {Humphreys},
  {Monard}, {Prieto}, \& {Walter}}]{bond09}
{Bond}, H.~E., {Bedin}, L.~R., {Bonanos}, A.~Z., {Humphreys}, R.~M., {Monard},
  L.~A.~G.~B., {Prieto}, J.~L., \& {Walter}, F.~M. 2009, \apjl, 695, L154

\bibitem[{{Botticella} {et~al.}(2009){Botticella}, {Pastorello}, {Smartt},
  {Meikle}, {Benetti}, {Kotak}, {Cappellaro}, {Crockett}, {Mattila}, {Sereno},
  {Patat}, {Tsvetkov}, {van Loon}, {Abraham}, {Agnoletto}, {Arbour}, {Benn},
  {di Rico}, {Elias-Rosa}, {Gorshanov}, {Harutyunyan}, {Hunter}, {Lorenzi},
  {Keenan}, {Maguire}, {Mendez}, {Mobberley}, {Navasardyan}, {Ries},
  {Stanishev}, {Taubenberger}, {Trundle}, {Turatto}, \& {Volkov}}]{bot09}
{Botticella}, M.~T. {et~al.} 2009, \mnras, 398, 1041

\bibitem[{{Bruenn} {et~al.}(2013){Bruenn}, {Mezzacappa}, {Hix}, {Lentz},
  {Bronson Messer}, {Lingerfelt}, {Blondin}, {Endeve}, {Marronetti}, \&
  {Yakunin}}]{bru13}
{Bruenn}, S.~W. {et~al.} 2013, \apjl, 767, L6

\bibitem[{{Burrows}(2013)}]{bur13}
{Burrows}, A. 2013, Reviews of Modern Physics, 85, 245

\bibitem[{{Burrows} {et~al.}(2007){Burrows}, {Dessart}, \& {Livne}}]{bur07}
{Burrows}, A., {Dessart}, L., \& {Livne}, E. 2007, in American Institute of
  Physics Conference Series, Vol. 937, Supernova 1987A: 20 Years After:
  Supernovae and Gamma-Ray Bursters, ed. S.~{Immler}, K.~{Weiler}, \&
  R.~{McCray}, 370--380

\bibitem[{{Duncan}(1921)}]{dun21}
{Duncan}, J.~C. 1921, Proceedings of the National Academy of Science, 7, 179

\bibitem[{{Dwek}(1983)}]{dwe83}
{Dwek}, E. 1983, \apj, 274, 175

\bibitem[{{Eastman} {et~al.}(1994){Eastman}, {Woosley}, {Weaver}, \&
  {Pinto}}]{eas94}
{Eastman}, R.~G., {Woosley}, S.~E., {Weaver}, T.~A., \& {Pinto}, P.~A. 1994,
  \apj, 430, 300

\bibitem[{{Fesen} {et~al.}(1997){Fesen}, {Shull}, \& {Hurford}}]{fes97}
{Fesen}, R.~A., {Shull}, J.~M., \& {Hurford}, A.~P. 1997, \aj, 113, 354

\bibitem[{{Frail} {et~al.}(1995){Frail}, {Kassim}, {Cornwell}, \&
  {Goss}}]{fra95}
{Frail}, D.~A., {Kassim}, N.~E., {Cornwell}, T.~J., \& {Goss}, W.~M. 1995,
  \apjl, 454, L129+

\bibitem[{{Hester}(2008)}]{hes08}
{Hester}, J.~J. 2008, \araa, 46, 127

\bibitem[{{Hoffman} {et~al.}(2008){Hoffman}, {M{\"u}ller}, \& {Janka}}]{hof08}
{Hoffman}, R.~D., {M{\"u}ller}, B., \& {Janka}, H. 2008, \apjl, 676, L127

\bibitem[{{Janka} {et~al.}(2008){Janka}, {M{\"u}ller}, {Kitaura}, \&
  {Buras}}]{jan08}
{Janka}, H., {M{\"u}ller}, B., {Kitaura}, F.~S., \& {Buras}, R. 2008, \aap,
  485, 199

\bibitem[{{Janka} {et~al.}(2012){Janka}, {Hanke}, {Huedepohl}, {Marek},
  {Mueller}, \& {Obergaulinger}}]{jan12}
{Janka}, H.-T., {Hanke}, F., {Huedepohl}, L., {Marek}, A., {Mueller}, B., \&
  {Obergaulinger}, M. 2012, ArXiv e-prints, 1211.1378

\bibitem[{{Kitaura} {et~al.}(2006){Kitaura}, {Janka}, \& {Hillebrandt}}]{kit06}
{Kitaura}, F.~S., {Janka}, H., \& {Hillebrandt}, W. 2006, \aap, 450, 345

\bibitem[{{Kochanek}(2011)}]{koc11}
{Kochanek}, C.~S. 2011, \apj, 741, 37

\bibitem[{{Kotake} {et~al.}(2012){Kotake}, {Sumiyoshi}, {Yamada}, {Takiwaki},
  {Kuroda}, {Suwa}, \& {Nagakura}}]{kot12}
{Kotake}, K., {Sumiyoshi}, K., {Yamada}, S., {Takiwaki}, T., {Kuroda}, T.,
  {Suwa}, Y., \& {Nagakura}, H. 2012, Progress of Theoretical and Experimental
  Physics, 2012, 301

\bibitem[{{Langer}(2012)}]{lan12}
{Langer}, N. 2012, \araa, 50, 107

\bibitem[{{Litvinova} \& {Nadezhin}(1985)}]{lit85}
{Litvinova}, I.~Y., \& {Nadezhin}, D.~K. 1985, Soviet Astronomy Letters, 11,
  145

\bibitem[{{Lundmark}(1921)}]{lun21}
{Lundmark}, K. 1921, \pasp, 33, 225

\bibitem[{{MacAlpine} \& {Satterfield}(2008)}]{mac08}
{MacAlpine}, G.~M., \& {Satterfield}, T.~J. 2008, \aj, 136, 2152

\bibitem[{{Matzner} \& {McKee}(1999)}]{mat99}
{Matzner}, C.~D., \& {McKee}, C.~F. 1999, \apj, 510, 379

\bibitem[{{Miller}(1973)}]{mil73}
{Miller}, J.~S. 1973, \apjl, 180, L83+

\bibitem[{{Miyaji} \& {Nomoto}(1987)}]{miy87}
{Miyaji}, S., \& {Nomoto}, K. 1987, \apj, 318, 307

\bibitem[{{Miyaji} {et~al.}(1980){Miyaji}, {Nomoto}, {Yokoi}, \&
  {Sugimoto}}]{miy80}
{Miyaji}, S., {Nomoto}, K., {Yokoi}, K., \& {Sugimoto}, D. 1980, \pasj, 32, 303

\bibitem[{{Nomoto}(1984)}]{nom84}
{Nomoto}, K. 1984, \apj, 277, 791

\bibitem[{{Nomoto}(1987)}]{nom87}
------. 1987, \apj, 322, 206

\bibitem[{{Nomoto} {et~al.}(1982){Nomoto}, {Sugimoto}, {Sparks}, {Fesen},
  {Gull}, \& {Miyaji}}]{nom82crab}
{Nomoto}, K., {Sugimoto}, D., {Sparks}, W.~M., {Fesen}, R.~A., {Gull}, T.~R.,
  \& {Miyaji}, S. 1982, \nat, 299, 803

\bibitem[{{Pearce} \& {Mayes}(1986)}]{pea86}
{Pearce}, G., \& {Mayes}, A.~J. 1986, \aap, 155, 291

\bibitem[{{Popov}(1993)}]{pop93}
{Popov}, D.~V. 1993, \apj, 414, 712

\bibitem[{{Prieto} {et~al.}(2008){Prieto}, {Kistler}, {Thompson}, {Y{\"u}ksel},
  {Kochanek}, {Stanek}, {Beacom}, {Martini}, {Pasquali}, \& {Bechtold}}]{pri08}
{Prieto}, J.~L. {et~al.} 2008, \apjl, 681, L9

\bibitem[{{Pskovskii}(1985)}]{psk85}
{Pskovskii}, I.~P. 1985, {Novye i sverkhnovye zvezdy}, ed. {Pskovskii, I.~P.}

\bibitem[{{Rudie} {et~al.}(2008){Rudie}, {Fesen}, \& {Yamada}}]{rud08}
{Rudie}, G.~C., {Fesen}, R.~A., \& {Yamada}, T. 2008, \mnras, 384, 1200

\bibitem[{{Sahu} {et~al.}(2006){Sahu}, {Anupama}, {Srividya}, \&
  {Muneer}}]{sah06}
{Sahu}, D.~K., {Anupama}, G.~C., {Srividya}, S., \& {Muneer}, S. 2006, \mnras,
  372, 1315

\bibitem[{{Saio} {et~al.}(1988){Saio}, {Nomoto}, \& {Kato}}]{sai88}
{Saio}, H., {Nomoto}, K., \& {Kato}, M. 1988, \apj, 331, 388

\bibitem[{{Siess}(2007)}]{sie07}
{Siess}, L. 2007, \aap, 476, 893

\bibitem[{{Smith}(2013)}]{smi13}
{Smith}, N. 2013, ArXiv e-prints, 1304.0689

\bibitem[{{Sollerman} {et~al.}(2001){Sollerman}, {Kozma}, \&
  {Lundqvist}}]{sol01}
{Sollerman}, J., {Kozma}, C., \& {Lundqvist}, P. 2001, \aap, 366, 197

\bibitem[{{Stephenson} \& {Green}(2002)}]{ste02}
{Stephenson}, F.~R., \& {Green}, D.~A. 2002, Historical supernovae and their
  remnants, by F.~Richard Stephenson and David A.~Green.~International series
  in astronomy and astrophysics, vol.~5.~Oxford: Clarendon Press, 2002, ISBN
  0198507666, 5

\bibitem[{{Szczygie{\l}} {et~al.}(2012){Szczygie{\l}}, {Kochanek}, \&
  {Dai}}]{szc12}
{Szczygie{\l}}, D.~M., {Kochanek}, C.~S., \& {Dai}, X. 2012, \apj, 760, 20

\bibitem[{{Tanaka} \& {Takahara}(2010)}]{tan10}
{Tanaka}, S.~J., \& {Takahara}, F. 2010, \apj, 715, 1248

\bibitem[{{Trimble}(1973)}]{tri73}
{Trimble}, V. 1973, \pasp, 85, 579

\bibitem[{{Vos}(1978)}]{vos78}
{Vos}, J.~J. 1978, Color Research \& Application, 3, 125

\bibitem[{{Wanajo} {et~al.}(2011){Wanajo}, {Janka}, \& {M{\"u}ller}}]{wanajo11}
{Wanajo}, S., {Janka}, H.-T., \& {M{\"u}ller}, B. 2011, \apjl, 726, L15

\bibitem[{{Wanajo} {et~al.}(2013){Wanajo}, {Janka}, \& {M{\"u}ller}}]{wanajo13}
------. 2013, \apjl, 767, L26

\bibitem[{{Wanajo} {et~al.}(2009){Wanajo}, {Nomoto}, {Janka}, {Kitaura}, \&
  {M{\"u}ller}}]{wanajo09}
{Wanajo}, S., {Nomoto}, K., {Janka}, H., {Kitaura}, F.~S., \& {M{\"u}ller}, B.
  2009, \apj, 695, 208

\bibitem[{{Zampieri} {et~al.}(2003){Zampieri}, {Pastorello}, {Turatto},
  {Cappellaro}, {Benetti}, {Altavilla}, {Mazzali}, \& {Hamuy}}]{zam03}
{Zampieri}, L., {Pastorello}, A., {Turatto}, M., {Cappellaro}, E., {Benetti},
  S., {Altavilla}, G., {Mazzali}, P., \& {Hamuy}, M. 2003, \mnras, 338, 711

\end{thebibliography}
\end{document}